# Probably faster multiplication of sparse polynomials*†

JORIS VAN DER HOEVEN

CNRS (UMI 3069, PIMS)
Department of Mathematics
Simon Fraser University
8888 University Drive
Burnaby, British Columbia
V5A 1S6, Canada

*Email:* vdhoeven@lix.polytechnique.fr

In this paper, we present a probabilistic algorithm to multiply two sparse polynomials almost as efficiently as two dense univariate polynomials with a result of approximately the same size. The algorithm depends on unproven heuristics that will be made precise. Non-heuristic versions that are a constant times slower are also presented.

## 1. INTRODUCTION

Let $P, Q \in \mathbb{Z}[x_1, \ldots, x_n]$ be polynomials that are represented in the usual way as linear combinations of power products. The problem of *sparse polynomial multiplication* is to compute the product $R = P Q$ in a way that is as efficient as possible in terms of the total bitsize of $P$, $Q$, and $R$ (and where we use a similar *sparse representation* for $R$ as for $P$ and $Q$).

For pedagogical reasons, we mainly restrict our attention to polynomials with integer coefficients. Together with polynomials with rational coefficients, this is indeed the most important case for practical implementations inside computer algebra systems. Nevertheless, it is not hard to adapt our techniques to coefficients in more general rings (some indications to that effect are given in section 5.2). Still for pedagogical reasons, we will carry out our complexity analysis in the RAM model [12]. We expect our algorithms to adapt to the Turing model [40], but more work will be needed to prove this and some of the constant factors might deteriorate.

For polynomials of modest size, naive algorithms are often most efficient. We refer to [4, 7, 11, 21, 32, 37, 38, 45] for implementation techniques that are efficient in practice. Various types of faster algorithms have been proposed for polynomials with special supports [18, 20, 26, 41].

Asymptotically fast methods for polynomials of large sizes usually rely on sparse interpolation. The seminal paper by Ben Or and Tiwari [5] triggered the development of many fast algorithms for the sparse interpolation of polynomial blackbox functions [1, 6,

*. This paper is part of a project that has received funding from the French "Agence de l'innovation de défense".

†. This article has been written using GNU $\text{T}_{\text{E}}\text{X}_{\text{MACS}}$ [25].

10, 22, 27–29, 31, 33, 39]. In this framework, the unknown polynomial $R$ is given through a blackbox function that can be evaluated at points in suitable extensions of the coefficient ring. We refer to [42] for a nice survey on sparse interpolation and other algorithms to compute with sparse polynomials. The present paper grew out of our recent preprint [24] with Grégoire Lecerf on this topic; the idea to "exploit colliding terms" in section 6.6 forms the starting point of this paper. We refer to [24] for more references and will also refer to that paper at several places for more background information.

The most efficient algorithms for sparse interpolation are mostly probabilistic. Here we note that it is usually easy to check that the result is correct with high probability: just evaluate both the blackbox function and its supposed interpolation at a random point and verify that both evaluations coincide. In this paper, all algorithms will be probabilistic, which is suitable for the practical purposes that we are interested in. The running times of our algorithms also rely on suitable heuristics that we will make precise.

Although the multiplication problem for sparse polynomials does not directly fit into the usual blackbox model, it does benefit from the techniques that have been developed for sparse interpolation. Practical algorithms along these lines have appeared in [9, 15, 21, 35]. Most algorithms operate in two phases: we first need to determine the exponents of the product $R$ and then its coefficients. The first phase is typically more expensive when the coefficients of $R$ are small, but it becomes cheap for large coefficients, due to the fact that we may first reduce $P, Q, R$ modulo a suitable prime. It is also customary to distinguish between the supersparse case in which the total degree of $R$ is allowed to become huge, the normally sparse case in which the total degree remains small, and the weakly sparse case when the total degree is very small with respect to the number of terms. In this paper, we mainly focus on the last asymptotic regime, which is most important for practical applications with a large number of variables.

In order to describe the complexity results, let us introduce some notations. Given a polynomial $P \in \mathbb{Z}[x_1, \ldots, x_n]$, we will write $d_P$ for its total degree, $t_P$ for its number of terms, $s_P$ for the number of powers $x_i^e$ with $e \geqslant 1$ that occur in its representation (identical powers being counted multiple times), and $|P|$ for the maximal absolute value of a coefficient. For instance, if $P = 3\, x_1^2 x_2^3 - 20\, x_2^3 x_3 x_4 + x_4^4$, then we have $d_P = 4$, $t_P = 3$, $s_P = 6$, and $|P| = 20$. For our multiplication problem $R = P\,Q$, the degree $d := d_R = d_P + d_Q$ of the result is easily determined, but we usually only assume an upper bound $T \geqslant t_R$ with $T = O(t_R)$ for its number of terms.

It is interesting to keep track of the dependency of our running times on logarithmic factors in certain parameters, but it is also convenient to ignore less important logarithmic and sublogarithmic factors. We do this by introducing the notation

$$f = O^\flat(g) \iff f = O(g\,(\log\,(s_P s_Q s_R |P|\,|Q|\,|R|))^{o(1)}\,(\log\,(d\,n))^{O(1)}).$$

We also wish to compare the cost of our algorithms with the cost of multiplying dense univariate polynomials of approximately the same size. Given integers $N, r > 1$, we therefore also introduce the following two complexities:

- $\mathsf{M}_N(r)$ stands for the cost of multiplying two non-zero polynomials in $\mathbb{Z}[u]/(u^r - 1)$ under the assumption that the product $R$ satisfies $|R| \leqslant N$.
- $\mathsf{M}'_N(r)$ stands for the cost of multiplying two polynomials in $(\mathbb{Z}/N\mathbb{Z})[u]/(u^r - 1)$.

We make the customary assumption that $\mathsf{M}_N(r)/r$ and $\mathsf{M}'_N(r)/r$ are non-decreasing as functions in $r$. By [16], one may take $\mathsf{M}_N(r) = O(r \log N \log(r \log N))$. If $r = O(N)$, then one also has $\mathsf{M}'_N(r) = O(\mathsf{M}_N(r))$, using Kronecker substitution [12].

One traditional approach for sparse polynomial multiplication is to evaluate $P$, $Q$, and $R$ at $2T$ points in a geometric progression $(p_1^k, p_2^k, \ldots, p_n^k)$ modulo a sufficiently large prime number $\Pi \asymp T p_n^d$, where $p_i$ stands for the $i$-th prime number. In combination with the tangent Graeffe method [24, sections 5 and 7.2] and fast smooth factoring (see, e.g. Lemma 5 below), this approach allows the exponents of $R$ to be computed in time

$$O^\flat(\mathsf{M}'_\Pi(T) \log \Pi). \tag{1}$$

The coefficients can be recovered using fast Vandermonde system solving, in time

$$O^\flat(\mathsf{M}'_N(t_R) \log t_R), \tag{2}$$

where $N = 2\, t_P\, t_Q\, |P|\, |Q| > 2\, |R| - 1$. In our case when $d$ is small, we usually have $\log \Pi = O^\flat(\log T)$, in which case (1) simplifies into $O^\flat(T\, (\log T)^3)$. The dependence of the complexity on $d$ can be reduced using techniques from [29], among others [22].

The main results of this paper are two faster probabilistic algorithms. The shorter running times rely on two heuristics **HE** and **HC** that will be detailed in section 4. For any $\tau > \tau_{\mathrm{crit}} \approx 0.407$, we show (Theorem 6) that the exponents of $R$ can be computed in time

$$6\, \tau\, \mathsf{M}'_\Pi(T) + O^\flat((s_P + s_Q + s_R) \log \Pi + (t_P + t_Q) \log N + t_R\, n), \tag{3}$$

where $N = t_P t_Q |P| |Q|$ and $\Pi \asymp T p_n^d$ is prime. This algorithm is probabilistic of Monte Carlo type. Based on numerical evidence in section 3, we conjecture that $\tau_{\mathrm{crit}} \approx 0.407265$. We also show (Theorem 4) that the coefficients may be computed in expected time

$$3\, \tau\, \mathsf{M}_N(t_R) + O^\flat((s_P + s_Q + s_R) \log t_R + (t_P + t_Q + t_R) \log N), \tag{4}$$

using a probabilistic algorithm of Las Vegas type. In practice, when $d$ is small and $n$ not too large with respect to $\log T$, the corrective terms in (3) are negligible and the cost reduces to $(6\, \tau + o(1))\, \mathsf{M}'_\Pi(T)$. Similarly, the cost (4) usually simplifies to $(3\, \tau + o(1))\, \mathsf{M}_N(t_R)$. If we also have $\Pi = o(N)$, then this means that the cost of the entire sparse multiplication becomes $(3\, \tau + o(1))\, \mathsf{M}_N(t)$. Here we note that $\mathsf{M}_N(t)$ also corresponds to the time needed to multiply two dense polynomials in $\mathbb{Z}[x]$, provided that the product $R$ satisfies $\deg R < t$ and $t\, |R| < N$.

The proof of these bounds relies on the evaluation of $R$ at three points of the form $(u^{\lambda_1}, \ldots, u^{\lambda_n})$ in algebras of the form $\mathbb{Z}[u] / (u^r - 1)$, where $r = \lfloor \tau t \rfloor$. If $\tau$ is sufficiently large (namely $\tau > \tau_{\mathrm{crit}}$ for some critical value) and we already know the exponents of $R$, then we show how to recover the coefficients with high probability. One interesting feature of our algorithm is that three evaluations are sufficient with high probability. A logarithmic number of evaluations is necessary when using the more obvious iterative approach for which every additional evaluation allows us to compute a constant fraction of the unknown coefficients (with high probability). Our algorithm is first explained with an example in section 2 and then in general in section 4.1. The probabilistic analysis is done in section 3. In section 4.2, we extend our approach to the computation of the exponents, using three additional evaluations. Section 5 is devoted to variants and extensions of our approach, and further remarks. The final section 6 contains slower but unconditional variants that do not depend on the heuristics **HE** and **HC**.

The present paper works out an idea that was first mentioned in [24, section 6.6], in the context of general sparse interpolation. The application to polynomial multiplication is particularly suitable because of the low amortized cost of blackbox evaluations. Random monomial transformations (also called randomized Kronecker substitution) were considered before in [3, 30]. The idea of using evaluations in (small) cyclic algebras has been used before in [2, 10, 35]. Our algorithms also exploit ideas from the technique of diversification [13, 31]. However, even though the present work boroughs a lot of techniques from these previous works, our final complexity bounds (3) and (4) improve on the previously known ones (see section 5.1 for variants in other asymptotic regimes). The approach to evaluate in cyclic algebras finally seems close to binning techniques that have recently been applied to compute sparse Fourier transforms [17, 34]; we plan to investigate this parallel in future work.

**Notes.** We released three versions of the present paper. In April 2022, we added section 6 with variants of our main algorithms that have the major "sales" advantage of being unconditional, but at the same time miss the more subtle charm of reducing the constant factor with respect to dense multiplication to almost one (or two for the exponents).

The present version integrates various minor corrections and adds a few attributions that were absent in the 2020 version. We have also been made aware of the fact that the "game of mystery balls" was already known in a different form [14]. The critical value $\tau_{\text{crit}}$ can actually be computed "exactly": $\tau_{\text{crit}} = \inf\{\alpha > 0 : \forall x \in (0,1), 1 - \mathrm{e}^{-x^2/\alpha} < x\}$.

Finally, we started an experimantal MATHEMAGIX implementation of our multiplication algorithms. We implemented the variant from section 5.1 over an FFT-field $\mathbb{F}_p$, where the exponents and $p$ all fit into machine double precision numbers. A typical multiplication of two multivariate sparse polynomials with $t_P \approx t_Q \approx 1.4 \cdot 10^6$ and $t_R \approx 1.1 \cdot 10^7$ takes about 8.2 seconds on an INTEL XEON processor running at 3.2 GHz. The "asymptotically dominant" cyclic multiplications in $\mathbb{F}_p[u]/(u^r - 1)$ take about 2.1 seconds, which is about 5 times the cost of a cyclic muliplication of length $t_R$, whereas $9\,\tau_{\text{crit}} \approx 3.6$ in (5). Note that the cyclic multiplications through DFTs benefit from heavier optimizations than the subdominant operations.

## 2. A GAME OF MYSTERY BALLS

Consider two sparse polynomials

$$\begin{aligned} P &= x y^5 + 3 x y^6 z - 2 x^8 y^{10} + x^{10} y^{14} z^3 \\ Q &= 2 + y z + 3 x^2 y^4 z^3. \end{aligned}$$

Their product $R = P Q$ is given by

$$\begin{aligned} R = {} & 3 x^{12} y^{18} z^6 + x^{10} y^{15} z^4 + 9 x^3 y^{10} z^4 + 3 x^3 y^9 z^3 - 4 x^{10} y^{14} z^3 + \\ & 3 x y^7 z^2 + 7 x y^6 z - 2 x^8 y^{11} z + 2 x y^5 - 4 x^8 y^{10}. \end{aligned}$$

Assume that the monomials $x^{12} y^8 z^6, x^{10} y^{15} z^4, \ldots$ of $R$ are known, but not the corresponding coefficients. Our aim is to determine $R$ through its evaluations at "points" of the form $(x, y, z) = (u^\alpha, u^\beta, u^\gamma)$ in $\mathbb{Z}[u]/(u^r - 1)$ for suitable $\alpha, \beta, \gamma \in \mathbb{N}$ and lengths $r$. These evaluations are obtained by evaluating $P$ and $Q$ at the same points and multiplying the results in $\mathbb{Z}[u]/(u^r - 1)$. In what follows, we will use three evaluation points.

Let us show how to turn the problem of computing the coefficients of $R$ into a "game of mystery balls". At the start of the game, we have one numbered ball for each term of $R$:

$$\begin{aligned} R = {} & \overbrace{3 x^{12} y^{18} z^6}^{①} + \overbrace{x^{10} y^{15} z^4}^{②} + \overbrace{9 x^3 y^{10} z^4}^{③} + \overbrace{3 x^3 y^9 z^3}^{④} + \overbrace{(-4)\, x^{10} y^{14} z^3}^{⑤} + \\ & \overbrace{3 x y^7 z^2}^{⑥} + \overbrace{7 x y^6 z}^{⑦} + \overbrace{(-2)\, x^8 y^{11} z}^{⑧} + \overbrace{2 x y^5}^{⑨} + \overbrace{(-4)\, x^8 y^{10}}^{⑩}. \end{aligned}$$

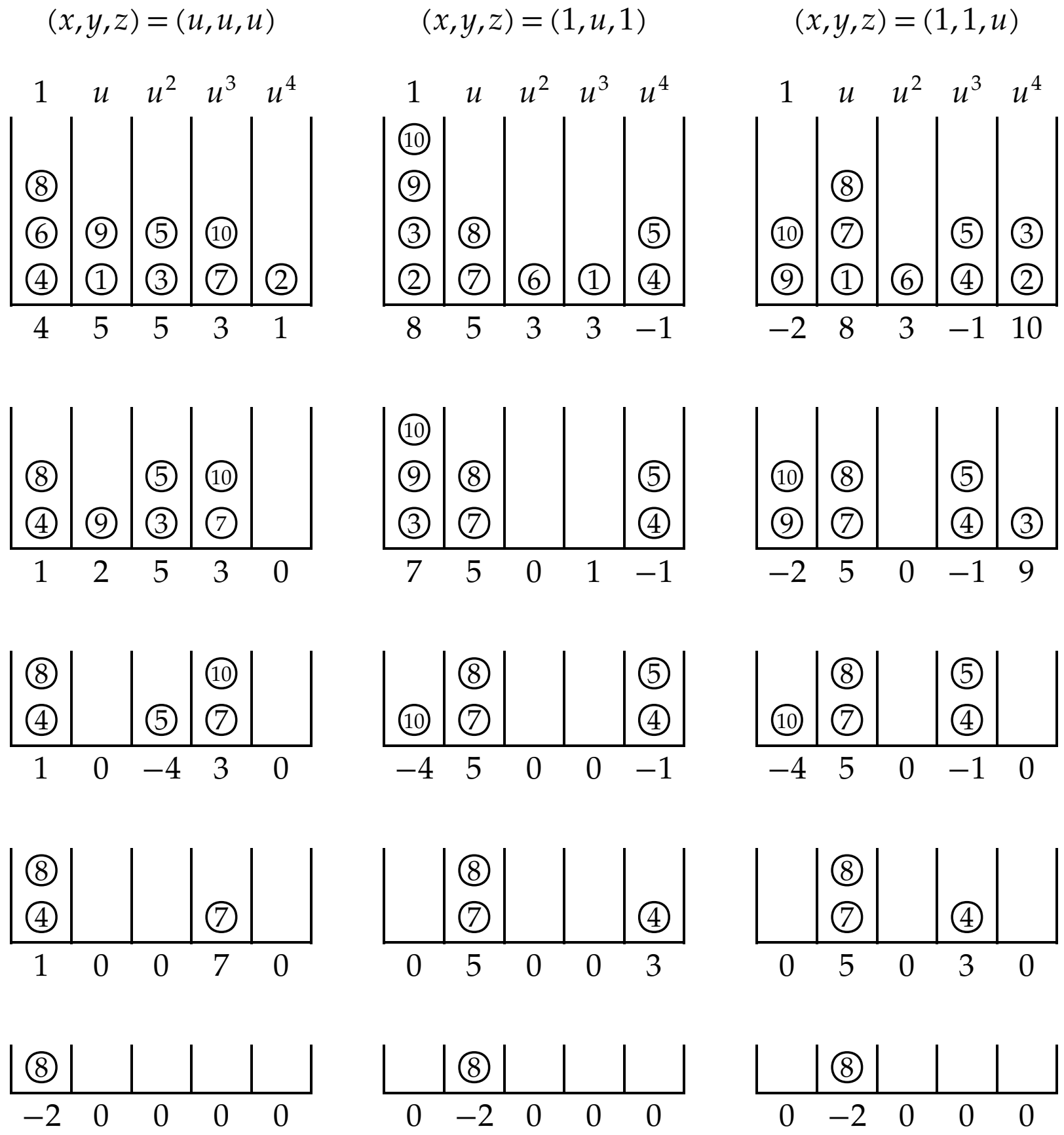

**Figure 1.** Playing the game of mystery balls. At every round, we remove the balls that ended up in a private box for at least one of the three throws.

For each ball, say ①, the corresponding "mystery coefficient" 3 needs to be determined (it might be hidden inside the ball), whereas the corresponding exponents $12, 18, 6$ are known (and stored in a table or painted on the ball). In fact, our game has three identical sets of balls, one for each of the three evaluation points. For each of these evaluation points, we also have a set of $r$ boxes, labeled by $1, u, u^2, \ldots, u^{r-1}$.

Now consider the evaluation of $R$ at a point as above, say at $(x,y,z) = (u,u,u)$ in the ring $\mathbb{Z}[u]/(u^5-1)$. Then each term $\kappa x^a y^b z^c$ evaluates to a term $\kappa u^e$ with $e \in \{0,1,2,3,4\}$ and $e = a+b+c$ modulo 5. In our game, we throw the corresponding ball into the box that is labeled by $u^e$. For instance, our first ball ① evaluates to $3u$ and goes into the box labeled by $u$. Our second ball ② evaluates to $1u^4$ and goes into the box labeled by $u^4$. Continuing this way, we obtain the upper left distribution in Figure 1. Now the complete evaluation of $R$ at $(x,y,z) = (u,u,u)$ in $\mathbb{Z}[u]/(u^5-1)$ gives

$$R(u,u,u) \;=\; 4 + 5u + 5u^2 + 3u^3 + u^4 \qquad (\operatorname{mod} u^5 - 1).$$

For each box, this means that we also know the sum of all coefficients hidden in the balls in that box. Indeed, in our example, the first box $u^0$ contains three balls ④, ⑥, and ⑧, with coefficients $3, 3$, and $-2$ that sum up to 4. In Figure 1, we indicated these sums below the boxes. In round one of our game, we actually took our chances three times, by using the three evaluation points $(u,u,u)$, $(1,u,1)$, and $(1,1,u)$ in $\mathbb{Z}[u]/(u^5-1)$, and throwing our balls accordingly. This corresponds to the top row in Figure 1.

Now we play our game as follows. If, in a certain round, a ball ends up alone in its box (we will also say that the ball has a private box), then the number below it coincides with the secret coefficient inside. At that point, we may remove the ball, as well as its copies from the two other throws, and update the numbers below accordingly. In round one of our running example, ball ② ends up in a box of its own for our first throw. Similarly, the balls ① and ⑥ both have private boxes for the second throw. Ball ⑥ also has a private box for the third throw. Removing the balls ①, ②, and ⑥ from the game, we obtain the second row in Figure 1. We also updated the numbers below the boxes: for every box, the number below it still coincides with the sum of the mystery coefficients inside the balls inside that box. Now that the balls ①, ②, and ⑥ have been removed, we observe that balls ③ and ⑨ have private boxes in their turn. We may thus determine their mystery coefficients and remove them from the game as well. This brings us to round three of our game and the third row in Figure 1. Going on like this, we win our game when all balls eventually get removed. We lose whenever there exists a round in which there are still some balls left, but all non-empty boxes contain at least two balls. In our example, we win after five rounds.

**Remark 1.** When implementing a computer program to play the game, we maintain a table that associates to each ball its exponents and the three boxes where it ended up for the three throws. Conversely, for each box, we maintain a linked list with all balls inside that box. We finally maintain a list with balls inside a private box, and which are about to be removed. In this way, if each box contains $O(1)$ balls, then the total amount of work that needs to be done in each round remains proportional to the number of balls that are removed instead of the total number of boxes. Consequently, the complete game can be played in linear time, with high probability.

**Remark 2.** Playing our game in several "rounds" is convenient for the probabilistic analysis in section 3 below. But the order in which balls in private boxes are removed actually does not matter, as long as we remove all balls in private boxes. Assume for instance that we win our game and that we replay it by removing balls in another order. Assume for contradiction that a ball $B$ does not get removed in our modified game and choose $B$ in such a way that the round $i$ in which it gets removed in the original game is minimal. Then we observe that all balls that were removed before round $i$ in the original game also get removed in the modified version, eventually. When this happens, $B$ is in a private box for one of the throws: a contradiction.

## 3. ON OUR PROBABILITY OF WINNING THE GAME

Have we been lucky in our example with 3 throws of 10 balls in 5 boxes? For the probabilistic analysis in this section, we will assume that our throws are random and independent. We will do our analysis for three throws, because this is best, although a similar analysis could be carried out for other numbers of throws. From now on, we will assume that we have $t$ balls and $r = \tau t$ boxes.

The experiment of throwing $t$ balls in $r$ boxes has widely been studied in the literature about hash tables [8, Chapter 9]. For a fixed ball, the probability that all other $t-1$ balls end up in another box is given by

$$p_1 = \left(1 - \frac{1}{r}\right)^{t-1} = \mathrm{e}^{(t-1)\log\left(1-\frac{1}{\tau t}\right)} = \mathrm{e}^{-\frac{1}{\tau}+O\left(\frac{1}{t}\right)} = \mathrm{e}^{-\frac{1}{\tau}} + O\left(\frac{1}{t}\right).$$

More generally, for any fixed $k \geqslant 1$, the probability that $k-1$ other balls end up in the same box and all the others in other boxes is given by

$$\begin{aligned} p_k &= \binom{t-1}{k-1} \frac{1}{r^{k-1}} \left(1-\frac{1}{r}\right)^{t-k} \\ &= \frac{t^{k-1}+O(t^{k-2})}{(k-1)!} \cdot \frac{1}{\tau^{k-1} t^{k-1}} \left(\mathrm{e}^{-\frac{1}{\tau}} + O\left(\frac{1}{t}\right)\right) \\ &= \frac{\mathrm{e}^{-\frac{1}{\tau}}}{(k-1)!\, \tau^{k-1}} + O\left(\frac{1}{t}\right). \end{aligned}$$

Stated otherwise, we may expect with high probability that approximately $p_1 t$ balls end up in a private box, approximately $p_2 t$ balls inside a box with one other ball, and so on.

This shows how we can expect our balls to be distributed in the first round of our game and at the limit when $t$ gets large. Assume more generally that we know the distribution $(p_{i,k})_{k \in \mathbb{N}}$ in round $i$ and let us show how to determine the distribution $(p_{i+1,k})_{k \in \mathbb{N}}$ for the next round. More precisely, assume that $p_{i,k}(t+O(1))$ is the expected number of balls in a box with $k$ balls in round $i$, where we start with

$$p_{1,k} = \frac{\mathrm{e}^{-\frac{1}{\tau}}}{(k-1)!\, \tau^{k-1}}.$$

Setting

$$\sigma_i = p_{i,1} + p_{i,2} + \cdots,$$

we notice that $1 = \sigma_1 > \sigma_2 > \sigma_3 > \cdots$, where $(\sigma_{i+1} - \sigma_i)\, t$ stands for the expected number of balls that are removed during round $i$.

Now let us focus on the first throw in round $i$ (the two other throws behave similarly, since they follow the same probability distribution). There are $p_{i,1} t$ balls that are in a private box for this throw. For each of the remaining balls, the probability $\pi_i$ that it ended up in a private box for at least one of the two other throws is

$$\pi_i = \left(2 - \frac{p_{i,1}}{\sigma_i}\right) \frac{p_{i,1}}{\sigma_i}.$$

The probability that a box with $k \geqslant 2$ balls becomes one with $j$ balls in the next round is therefore given by

$$\lambda_{j,k} = \binom{k}{j} \pi_i^{k-j} (1-\pi_i)^j.$$

For all $j \geqslant 1$, this yields

$$p_{i+1,j} = \sum_{k \geqslant \max(2,j)} \frac{j}{k} \lambda_{j,k} p_{i,k}.$$

If $\sigma_i$ tends to zero for large $i$ and $t$ gets large, then we win our game with high probability. If $\sigma_i$ tends to a limit $\ell \in (0,1)$, then we will probably lose and end up with approximately $\ell t$ balls that can't be removed (for each of the three throws).

| $p_{i,k}$ | $k=1$ | 2 | 3 | 4 | 5 | 6 | 7 | $\sigma_i$ |
|---|---|---|---|---|---|---|---|---|
| $i=1$ | 0.13534 | 0.27067 | 0.27067 | 0.18045 | 0.09022 | 0.03609 | 0.01203 | 1.00000 |
| 2 | 0.06643 | 0.25063 | 0.18738 | 0.09340 | 0.03491 | 0.01044 | 0.00260 | 0.64646 |
| 3 | 0.04567 | 0.21741 | 0.13085 | 0.05251 | 0.01580 | 0.00380 | 0.00076 | 0.46696 |
| 4 | 0.03690 | 0.18019 | 0.08828 | 0.02883 | 0.00706 | 0.00138 | 0.00023 | 0.34292 |
| 5 | 0.03234 | 0.13952 | 0.05443 | 0.01416 | 0.00276 | 0.00043 | 0.00006 | 0.24371 |
| 6 | 0.02869 | 0.09578 | 0.02811 | 0.00550 | 0.00081 | 0.00009 | 0.00001 | 0.15899 |
| 7 | 0.02330 | 0.05240 | 0.01033 | 0.00136 | 0.00013 | 0.00001 | 0.00000 | 0.08752 |
| 8 | 0.01428 | 0.01823 | 0.00193 | 0.00014 | 0.00001 | 0.00000 | 0.00000 | 0.03459 |
| 9 | 0.00442 | 0.00249 | 0.00009 | 0.00000 | 0.00000 | 0.00000 | 0.00000 | 0.00700 |
| 10 | 0.00030 | 0.00005 | 0.00000 | 0.00000 | 0.00000 | 0.00000 | 0.00000 | 0.00035 |
| 11 | 0.00000 | 0.00000 | 0.00000 | 0.00000 | 0.00000 | 0.00000 | 0.00000 | 0.00000 |

**Table 1.** The probability distributions $(p_{i,k})_{k\in\mathbb{N}}$ in rounds $i=1,\ldots,11$ for $\tau=1/2$.

We have not yet been able to fully describe the asymptotic behavior of the distribution $(p_{i,k})_{k\in\mathbb{N}}$ for $i\to\infty$, which follows a non-linear dynamics. Nevertheless, it is easy to compute reliable approximations for the coefficients $p_{i,k}$ using interval or ball arithmetic [19]; for this purpose, it suffices to replace each coefficient $p_{i,k}$ in the tail of the distribution (i.e. for large $k$) by the interval $[0,p_{1,k}]$. Tables 1, 2, and 3 show some numerical data that we computed in this way (the error in the numbers being at most $0.5\cdot 10^{-5}$). Our numerical experiments indicate that the "phase change" between winning and losing occurs at a critical value $\tau_{\text{crit}}$ with

$$0.407264<\tau_{\text{crit}}<0.407265.$$

Table 1 shows what happens for $\tau=1/2$: until the seventh round, a bit less than half of the balls get removed at every round. After round eight, the remaining balls are removed at an accelerated rate. For $\tau=1/3$, the distributions $(p_{i,k})_{k\in\mathbb{N}}$ numerically tend to a non-zero limit distribution $(p_{\infty,k})_{k\in\mathbb{N}}$ with $\sigma_\infty := p_{\infty,1}+p_{\infty,2}+\cdots\approx 0.78350$. In Table 3, we show some of the distributions in round ten, for $\tau$ near the critical point $\tau_{\text{crit}}$. We also computed an approximation of the limit distribution at the critical point itself.

| $p_{i,k}$ | $k=1$ | 2 | 3 | 4 | 5 | 6 | 7 | $\sigma_i$ |
|---|---|---|---|---|---|---|---|---|
| $i=1$ | 0.04979 | 0.14936 | 0.22404 | 0.22404 | 0.16803 | 0.10082 | 0.05041 | 1.00000 |
| 2 | 0.01520 | 0.16294 | 0.22068 | 0.19925 | 0.13493 | 0.07310 | 0.03300 | 0.85795 |
| 3 | 0.00579 | 0.16683 | 0.21802 | 0.18994 | 0.12410 | 0.06487 | 0.02826 | 0.81315 |
| 4 | 0.00238 | 0.16826 | 0.21676 | 0.18616 | 0.11991 | 0.06179 | 0.02584 | 0.79590 |
| 5 | 0.00101 | 0.16883 | 0.21620 | 0.18457 | 0.11818 | 0.06053 | 0.02584 | 0.78878 |
| 6 | 0.00043 | 0.16907 | 0.21596 | 0.18389 | 0.11744 | 0.06000 | 0.02555 | 0.78577 |
| 7 | 0.00019 | 0.16918 | 0.21585 | 0.18360 | 0.11713 | 0.05978 | 0.02542 | 0.78448 |
| 8 | 0.00008 | 0.16922 | 0.21581 | 0.18347 | 0.11699 | 0.05968 | 0.02535 | 0.78392 |
| 9 | 0.00003 | 0.16924 | 0.21579 | 0.18342 | 0.11693 | 0.05964 | 0.02535 | 0.78368 |
| 10 | 0.00001 | 0.16925 | 0.21578 | 0.18340 | 0.11691 | 0.05962 | 0.02534 | 0.78358 |
| 11 | 0.00001 | 0.16925 | 0.21577 | 0.18339 | 0.11690 | 0.05961 | 0.02533 | 0.78353 |
| 12 | 0.00000 | 0.16925 | 0.21577 | 0.18338 | 0.11690 | 0.05961 | 0.02533 | 0.78351 |

**Table 2.** The probability distributions $(p_{i,k})_{k\in\mathbb{N}}$ in rounds $i=1,\ldots,11$ for $\tau=1/3$.

| $p_{10,k}$ | $k=1$ | 2 | 3 | 4 | 5 | 6 | 7 | $\sigma_{10}$ |
|---|---|---|---|---|---|---|---|---|
| $\tau=0.333$ | 0.00001 | 0.16892 | 0.21573 | 0.18368 | 0.11729 | 0.05992 | 0.02551 | 0.78447 |
| $\tau=0.400$ | 0.00190 | 0.20779 | 0.16840 | 0.09099 | 0.03687 | 0.01195 | 0.00323 | 0.52207 |
| $\tau=0.405$ | 0.00258 | 0.20592 | 0.15856 | 0.08140 | 0.03134 | 0.00965 | 0.00248 | 0.49260 |
| $\tau=0.407$ | 0.00290 | 0.20479 | 0.15429 | 0.07750 | 0.02919 | 0.00880 | 0.00221 | 0.48027 |
| $\tau=0.410$ | 0.00346 | 0.20266 | 0.14752 | 0.07159 | 0.02606 | 0.00759 | 0.00184 | 0.46118 |
| $\tau=0.420$ | 0.00601 | 0.19090 | 0.12180 | 0.05181 | 0.01653 | 0.00422 | 0.00090 | 0.39235 |
| $\tau=0.450$ | 0.01841 | 0.09767 | 0.03137 | 0.00672 | 0.00108 | 0.00014 | 0.00001 | 0.15541 |
| $\tau=0.500$ | 0.00030 | 0.00005 | 0.00000 | 0.00000 | 0.00000 | 0.00000 | 0.00000 | 0.00035 |
| $p_{10000,k}$, $\tau_{\text{crit}}$ | 0.00000 | 0.18338 | 0.11551 | 0.04851 | 0.01528 | 0.00385 | 0.00081 | 0.36751 |

**Table 3.** The probability distributions $(p_{10,k})_{k\in\mathbb{N}}$ in round 10 for various $\tau$ as well as an approximation of the limit distribution for $\tau=\tau_{\text{crit}}$, by taking the distribution in round 10000 for $\tau\approx 0.407264\approx\tau_{\text{crit}}$.

For $\tau<\tau_{\text{crit}}$, we note that reliable numerical computations can be turned into an actual proof that we lose with high probability. Indeed, assume that $\varepsilon:=p_{i,1}=o(1)$ gets very small for some $i$, whereas $\sigma_i^{-1}=O(1)$ remains bounded (for instance, in Table 2, we have $p_{i,1}<10^{-5}$ and $\sigma_i=0.78351$ for $i=12$). Then $\pi_i=(2+O(\varepsilon))\frac{\varepsilon}{\sigma_i}=O(\varepsilon)$ also gets very small and

$$\begin{aligned} p_{i+1,1} &= \sum_{k\geqslant 2} \pi_i^{k-1}(1-\pi_i)\,p_{i,k} \\ &= p_{i,2}\pi_i+O(\varepsilon^2) \\ &= 2\frac{p_{i,2}}{\sigma_i}\varepsilon+O(\varepsilon^2) \\ p_{i+1,k} &= p_{i,k}+O(\varepsilon), \qquad k\geqslant 2. \end{aligned}$$

If $2p_{i,2}$ happens to be $\nu$ times smaller than $\sigma_i$ for some fixed $\nu>1$ (for instance, in Table 2, we actually have $4p_{i,2}<\sigma_i$ for $i=12$), then a standard contracting ball type argument can be used to prove that $p_{i',1}$ decreases to zero with geometric speed for $i'\geqslant i$, while $\sigma_{i'}$ remains bounded away from zero.

Conversely, given $\tau>\tau_{\text{crit}}$, it seems harder to prove in a similar way that we win. Nevertheless, for any $\varepsilon>0$, reliable computations easily allow us to determine some round $i$ with $\sigma_i<\varepsilon$. For instance, for $\tau=1/2$ and $\varepsilon=10^{-5}$, we may take $i=11$. This is good enough for practical purposes: for $t=10000\ll\varepsilon^{-1}$, it means that we indeed win with high probability (and even if we do not win, then we may rerun the algorithm for the missing terms: see [1] and section 6 below). Here we also note that a large number of rounds are generally necessary to win when $\tau$ approaches the critical value $\tau_{\text{crit}}$. For instance, for $\tau=0.42$, we need to wait until round $i=29$ to get $\sigma_i<10^{-5}$. Nevertheless, after a certain number of rounds, it seems that $\sigma_i$ always converges to zero with superlinear speed.

In Table 4, we conclude with the result of a computer simulation in which we played our game with $t=100000$ and $\tau=1/2$. As one can see, the results are close to the theoretically expected ones from Table 1.

| $N_{i,k}$ | $k=1$ | 2 | 3 | 4 | 5 | 6 | 7 | $\Sigma_i$ |
|---|---|---|---|---|---|---|---|---|
| $i=1$ | 13438 | 27132 | 27027 | 18072 | 9075 | 3516 | 1260 | 100000 |
| 2 | 6577 | 25344 | 18576 | 9360 | 3575 | 990 | 259 | 64793 |
| 3 | 4524 | 22004 | 13149 | 5196 | 1535 | 354 | 84 | 46854 |
| 4 | 3649 | 18310 | 8904 | 2808 | 665 | 156 | 14 | 34506 |
| 5 | 3247 | 14190 | 5406 | 1464 | 295 | 24 | 0 | 24626 |
| 6 | 2849 | 9892 | 2823 | 556 | 65 | 6 | 0 | 16191 |
| 7 | 2327 | 5522 | 1071 | 124 | 15 | 0 | 0 | 9059 |
| 8 | 1501 | 1946 | 225 | 12 | 5 | 0 | 0 | 3689 |
| 9 | 487 | 256 | 18 | 0 | 0 | 0 | 0 | 761 |
| 10 | 34 | 6 | 0 | 0 | 0 | 0 | 0 | 40 |
| 11 | 0 | 0 | 0 | 0 | 0 | 0 | 0 | 0 |

**Table 4.** Carrying out a random simulation of our game for $t=100000$ and $\tau=1/2$. The table shows the number $N_{i,k}$ of balls that are in a box with $k$ balls in round $i$ (for the first throw). The last column also shows the sum $\Sigma_i = N_{i,1} + N_{i,2} + \cdots$.

**Remark 3.** Our game of mystery balls readily generalizes to different numbers of throws and different numbers of boxes for each throw. However, if our aim is to take the total number of boxes as small as possible, then computer experiments suggest that using three throws with the same number of boxes is optimal. For simplicity, we therefore restricted ourselves to this case, although it should be easy to extend our analysis to a more general setting.

## 4. MULTIPLYING SPARSE POLYNOMIALS

Let us now turn to the general problem of multiplying sparse polynomials. We will focus on the multiplication $R=PQ$ of integer polynomials $P,Q\in\mathbb{Z}[x_1,\ldots,x_n]$ in at least two variables. We define $d$ to be the total degree of $R$ and assume that we have a bound $T$ for the number of terms of $R$.

As explained in the introduction, we proceed in two phases: we first determine the exponents of the unknown product $R$. This part is probabilistic of Monte Carlo type, where we tolerate a rate of failure $\varepsilon>0$ that is fixed in advance. In the second phase, we determine the unknown coefficients, using a probabilistic algorithm of Las Vegas type. In this section, we start with the second phase, which has already been explained on an example in section 2.

### 4.1. Determination of the coefficients

Assume that our product $R=PQ$ has $t$ terms, so that $R=c_1 x^{e_1}+\cdots+c_t x^{e_t}$ for certain $c_1,\ldots,c_t\in\mathbb{Z}$ and $e_1,\ldots,e_t\in\mathbb{N}^n$, where $x^{e_i}=x_1^{e_{i,1}}\cdots x_n^{e_{i,n}}$ for $i=1,\ldots,t$. It is obvious how to generalize the algorithm from section 2 to this case: for some fixed $\tau>\tau_{\text{crit}}$, we distribute "our $t$ balls" over $r=\lfloor\tau t\rfloor$ boxes, through the evaluation of $R$ at three points $(u^{\lambda_{i,1}},\ldots,u^{\lambda_{i,n}})$ in $\mathbb{Z}[u]/(u^r-1)$ for $i=1,2,3$ and $\lambda_1,\lambda_2,\lambda_3\in\mathbb{Z}^n$. The vectors $\lambda_1,\lambda_2,\lambda_3$ are essentially chosen at random, but it is convenient to take them pairwise non-collinear modulo $r$, so as to avoid any "useless throws". We assume the following heuristic:

**HE.** For random vectors $\lambda_1,\lambda_2,\lambda_3$ as above and each of the three throws, the balls are distributed over the boxes in a uniformly random way, and the three distributions are independent.

**Algorithm 1**
**Input:** $P, Q \in \mathbb{Z}[x_1, \ldots, x_n]$ and $e_1, \ldots, e_t \in \mathbb{N}^n$ with $PQ \in \mathbb{Z}x^{e_1} + \cdots + \mathbb{Z}x^{e_t}$
**Output:** the product $PQ$, with high probability (Las Vegas), or "failed"
**Assume:** $n \geqslant 2$ and $t \geqslant 6$

1. For a fixed $\tau > \tau_{\text{crit}}$, let $r := \lfloor \tau t \rfloor$.
2. Let $\lambda_1, \lambda_2, \lambda_3 \in \{0, \ldots, r-1\}^n$ be random vectors that are pairwise non-collinear modulo $r$.
3. Compute $P_i = P(u^{\lambda_{i,1}}, \ldots, u^{\lambda_{i,n}})$ and $Q_i = Q(u^{\lambda_{i,1}}, \ldots, u^{\lambda_{i,n}})$ in $\mathbb{Z}[u]/(u^r - 1)$, for $i = 1, 2, 3$.
4. Multiply $R_i := P_i Q_i$ in $\mathbb{Z}[u]/(u^r - 1)$, for $i = 1, 2, 3$.
5. Let $J := \{1, \ldots, t\}$.
6. While there exist $i \in \{1, 2, 3\}$ and $j \in J$ with $\lambda_i \cdot e_j \neq \lambda_i \cdot e_{j'}$ for all $j' \in J \setminus \{j\}$, do
   a. Let $c_j$ be the coefficient of $u^{\lambda_i \cdot e_j}$ in $R_i$.
   b. For $i' \in \{1, 2, 3\}$, replace $R_{i'} := R_{i'} - c_j u^{\lambda_{i'} \cdot e_j}$.
   c. Replace $J := J \setminus \{j\}$.
7. Return $c_1 x^{e_1} + \cdots + c_t x^{e_t}$ if $J = \emptyset$ and "failed" otherwise.

We then play our game of mystery balls as usual, which yields the coefficients $c_1, \ldots, c_t$ if we win and only a subset of these coefficients if we lose. In view of Remark 2, this leads to Algorithm 1.

THEOREM 4. *Assume heuristic* **HE**. *Let $\tau > \tau_{\text{crit}}$ with $\tau \leqslant 1$ and $N = t_P t_Q |P| |Q|$. Then Algorithm 1 is correct and runs in time*

$$3\tau \mathsf{M}_N(t_R) + O^\flat((s_P + s_Q + s_R) \log t_R + (t_P + t_Q + t_R) \log N).$$

**Proof.** The correctness of the algorithm has been explained in section 2. As to the running time, we first note that none of the integer coefficients encountered during our computation exceeds $N$ in absolute value. Now steps 1 and 2 have negligible cost and the running time for the remaining steps is as follows:

- In step 3, for every term $c x^e$ in $P$ or $Q$ and every $i \in \{1, 2, 3\}$, we first have to compute $\lambda_i \cdot e = \lambda_{i,1} e_1 + \cdots + \lambda_{i,n} e_n$ modulo $r$. Since we only have to accumulate $\lambda_{i,j} e_j$ when $e_i \neq 0$, this can be done in time $O^\flat((s_P + s_Q) \log r)$. We next have to add the coefficients of all terms that end up in the same box, which amounts to $O(t_P + t_Q)$ additions of cost $O((t_P + t_Q) \log N)$.
- In step 4, we do three multiplications of cost $\mathsf{M}_N(r)$ each. Since $\mathsf{M}_N(r)/r$ is nondecreasing, the cost of these multiplications is bounded by $3\tau \mathsf{M}_N(t_R)$.
- In steps 5 and 6, we play our game of mystery balls, where $J$ stands for the set of ball that are still in play. Whenever ball $j$ ends up in a private box for throw number $i$, we have to determine where this ball landed for the other throws, by computing $\lambda_{i'} \cdot e_j$ modulo $r$ for $i' \neq i$. We then have to update the numbers below the corresponding boxes, which corresponds to setting $R_{i'} := R_{i'} - c_j u^{\lambda_{i'} \cdot e_j}$ in step 6b. Since this eventually has to be done for each of the $t$ balls, step 6 takes $O(s_R \log r + t_R \log N)$ bit-operations, using a similar analysis as for step 3.

Let us finally investigate bookkeeping costs that are implicit in our description of the algorithm. Above all, we have to maintain the linked lists with balls inside each box (see Remark 1). This can be done in time

$$O\left(\sum_{k\geqslant 1} k p_{1,k} T\right) = O\left(\sum_{k\geqslant 1} \frac{k}{(k-1)!\,\tau^{k-1}} \mathrm{e}^{-1/\tau} T\right) = O(T).$$

The other implicit costs to maintain various tables are also bounded by $O(T)$. □

## 4.2. Determination of the exponents

In [24], we surveyed several strategies for computing the exponents of the product $R$. Most of the approaches from sections 4, 6, and 7 of that paper can be adapted to the present setting. We will focus on a probabilistic strategy that we expect to be one of the most efficient ones for practical purposes (a few variants will be discussed in section 5).

For $i=1,2,\ldots$, let $p_i$ be the $i$-th prime number and let $B=p_n^d$. We let $\Pi$ be a fixed prime number with $\Pi\geqslant BT/\varepsilon$ (for practical implementations, we may also take $\Pi$ to be a product of prime numbers that fit into machine words, and use multi-modular arithmetic to compute modulo $\Pi$). For some fixed $\tau>\tau_{\text{crit}}$, we again use $r=\lfloor \tau T\rfloor$ boxes, and evaluate $P,Q,R$ over the ring $(\mathbb{Z}/\Pi\mathbb{Z})[u]/(u^r-1)$. This time, we use six evaluation points of the form $(u^{\lambda_{i,1}},\ldots,u^{\lambda_{i,n}})$ and $(p_1 u^{\lambda_{i,1}},\ldots,p_n u^{\lambda_{i,n}})$ for $i=1,2,3$, where the $\lambda_1,\lambda_2,\lambda_3$ are chosen at random and pairwise non-collinear modulo $r$.

Now consider a term $c x_1^{k_1}\cdots x_n^{k_n}$ of $R$. Its evaluation at $(u^{\lambda_{i,1}},\ldots,u^{\lambda_{i,n}})$ is $c u^e$, where $e=\lambda_{i,1}k_1+\cdots+\lambda_{i,n}k_n$ modulo $r$. Meanwhile, its evaluation at $(p_1 u^{\lambda_{i,1}},\ldots,p_n u^{\lambda_{i,n}})$ is $\tilde{c} u^e$ with $\tilde{c}=p_1^{k_1}\cdots p_n^{k_n} c$. If there is no other term that evaluates to an expression of the form $c' u^e$ at $(u^{\lambda_{i,1}},\ldots,u^{\lambda_{i,n}})$, then the same holds for the evaluation at $(p_1 u^{\lambda_{i,1}},\ldots,p_n u^{\lambda_{i,n}})$. Consequently, the unique representative $q\in\{0,\ldots,\Pi-1\}$ of the quotient $\tilde{c}/c$ in $\mathbb{Z}/\Pi\mathbb{Z}$ coincides with $p_1^{k_1}\cdots p_n^{k_n}$ (the quotient is well defined with high probability). From this, we can determine the exponents $k_1,\ldots,k_n$ by factoring $q$. As additional safeguards, we also check that $q\leqslant B$, that all prime factors of $q$ are in $\{p_1,\ldots,p_n\}$, and that $e=\lambda_{i,1}k_1+\cdots+\lambda_{i,n}k_n$ modulo $r$.

Conversely, assume that there are at least two terms of $R$ that evaluate to an expression of the form $c' u^e$ at $(u^{\lambda_{i,1}},\ldots,u^{\lambda_{i,n}})$. Let $c$ and $\tilde{c}$ now be the coefficients of $u^e$ in $R(u^{\lambda_{i,1}},\ldots,u^{\lambda_{i,n}})$ and $R(p_1 u^{\lambda_{i,1}},\ldots,p_n u^{\lambda_{i,n}})$. Then the quotient $\tilde{c}/c$ is essentially a random element in $\mathbb{Z}/\Pi\mathbb{Z}$ (see the heuristic **HC** below for more details), so its unique representative in $\{0,\ldots,\Pi-1\}$ is higher than $B$ with probability $1-\varepsilon/T$. This allows our algorithm to detect that we are dealing with colliding terms; for the $O(T)$ quotients that we need to consider the probability of failure becomes $1-(1-\varepsilon/T)^{O(T)}=O(\varepsilon)$.

Using the above technique, the three additional evaluations at $(p_1 u^{\lambda_{i,1}},\ldots,p_n u^{\lambda_{i,n}})_{i=1,2,3}$, allow us to determine which balls end up in a private box in our game of mystery balls. Moreover, since we can determine the corresponding exponents, we can also find where these balls landed for the two other throws. This allows us to play our game modulo minor modifications. Besides maintaining the numbers below the boxes for the first three throws, we also maintain the corresponding numbers for the evaluations at $(p_1 u^{\lambda_{i,1}},\ldots,p_n u^{\lambda_{i,n}})_{i=1,2,3}$. For every round of the game, the same technique then allows us to determine which balls have private boxes, and iterate.

**Algorithm 2**
**Input:** $P,Q\in\mathbb{Z}[x_1,\ldots,x_n]$ and a bound $T\geqslant t_{PQ}$
**Output:** the exponents of the product $PQ$, with high probability (Monte Carlo)
**Assume:** $n\geqslant 2$, $t\geqslant 6$, and a prime number $\Pi\geqslant\lceil p_n^d T/\varepsilon\rceil$ is known

1. For a fixed $\tau>\tau_{\text{crit}}$, let $r:=\lfloor\tau t\rfloor$, $B:=p_n^d$, and $\Pi\geqslant\lceil BT/\varepsilon\rceil$ prime.
2. Let $\lambda_1,\lambda_2,\lambda_3\in\{0,\ldots,r-1\}^n$ be random vectors that are pairwise non-collinear modulo $r$.
3. Compute the reductions $\bar{P},\bar{Q}\in(\mathbb{Z}/\Pi\mathbb{Z})[x_1,\ldots,x_n]$ of $P,Q$ modulo $\Pi$.
4. Let $v_1,\ldots,v_n$ be random invertible elements in $\mathbb{Z}/\Pi\mathbb{Z}$.
   Replace $\bar{P}:=\bar{P}(v_1x_1,\ldots,v_nx_n)$ and $\bar{Q}:=\bar{Q}(v_1x_1,\ldots,v_nx_n)$.
   Compute $\tilde{P}:=\bar{P}(p_1x_1,\ldots,p_nx_n)$ and $\tilde{Q}:=\bar{Q}(p_1x_1,\ldots,p_nx_n)$ in $(\mathbb{Z}/\Pi\mathbb{Z})[x_1,\ldots,x_n]$.
5. Compute the evaluations $\bar{P}_i$, $\bar{Q}_i$, $\tilde{P}_i$, $\tilde{Q}_i$ of $\bar{P}$, $\bar{Q}$, $\tilde{P}$, $\tilde{Q}$ at $(u^{\lambda_{i,1}},\ldots,u^{\lambda_{i,n}})$ for $i=1,2,3$.
6. Multiply $\bar{R}_i:=\bar{P}_i\bar{Q}_i$ and $\tilde{R}_i:=\tilde{P}_i\tilde{Q}_i$ in $(\mathbb{Z}/\Pi\mathbb{Z})[u]/(u^r-1)$, for $i=1,2,3$.
7. Let $\Omega:=\{(i,e)\in\{1,2,3\}\times\{1,\ldots,r\}:\bar{R}_{i,e}\neq 0\}$ and $E:=\emptyset$.
   For all $(i,e)\in\Omega$, compute the preimage $q_{i,e}\in\{0,\ldots,\Pi-1\}$ of $\tilde{R}_{i,e}/\bar{R}_{i,e}$.
   For all $(i,e)\in\Omega$, try factoring $q_{i,e}=p_1^{k_1}\cdots p_n^{k_n}$ whenever $q_{i,e}\leqslant B$.
8. While there exist $(i,e)\in\Omega$ with $q_{i,e}=p_1^{k_1}\cdots p_n^{k_n}$, $k_1+\cdots+k_n\leqslant d$, and $k\cdot\lambda_i=e$, do
   a. Let $c$ and $\tilde{c}$ be the coefficients of $u^{\lambda_i\cdot k}$ in $\bar{R}_i$ and $\tilde{R}_i$.
   b. For $i'\in\{1,2,3\}$, replace $\bar{R}_{i'}:=\bar{R}_{i'}-c\,u^{\lambda_{i'}\cdot k}$ and $\tilde{R}_{i'}:=\tilde{R}_{i'}-\tilde{c}\,u^{\lambda_{i'}\cdot k}$.
   c. Also update $q_{i',e'}:=\tilde{R}_{i,e'}/\bar{R}_{i,e'}$ and its factorization if $\bar{R}_{i,e'}\neq 0$ for $e'=\lambda_{i'}\cdot k$.
   d. Replace $\Omega:=\Omega\setminus\{(i',\lambda_{i'}\cdot k):i'=1,2,3\}$ and $E:=E\cup\{k\}$.
9. Return $E$ if $\Omega=\emptyset$ and "failed" otherwise.

In our probabilistic analysis, it is important that the quotients $\tilde{c}/c$ are essentially random elements in $\mathbb{Z}/\Pi\mathbb{Z}$ in case of collisions. This assumption might fail when the coefficients of $P$ and $Q$ are special (e.g. either zero or one). Nevertheless, it becomes plausible after a change of variables $x_i\mapsto v_ix_i$, $i=1,\ldots,n$, where $v_1,\ldots,v_n$ are random invertible elements in $\mathbb{Z}/\Pi\mathbb{Z}$. Let us formulate this as our second heuristic:

**HC.** For random vectors $\lambda_1,\lambda_2,\lambda_3$ and after a random change of variables $x_i\mapsto v_ix_i$ $(i=1,\ldots,n)$, the quotients $\tilde{c}/c$ as above are uniformly distributed over $\mathbb{Z}/\Pi\mathbb{Z}$, and the distributions for different quotients are independent.

Random change of variables $x_i\mapsto v_ix_i$ are a useful known technique in sparse interpolation [31] and the way we use it here is related to "diversification" [13]. Algorithm 2 summarizes our method, while incorporating the random change of variables.

LEMMA 5. *Given a number $q\in\{0,\ldots,\Pi-1\}$ with $s$ distinct prime factors, it takes*

$$O^\flat(n+s\log\Pi)$$

*bit operations to determine the existence of a prime factorization $q=p_1^{k_1}\cdots p_n^{k_n}$ with $k_1+\cdots+k_n\leqslant d$, and to compute it in case there is.*

**Proof.** We first determine the indices $k$ with $k_j \neq 0$ using a divide and conquer technique. At the top level, we start with the remainder $g$ of the division of $q$ by $p_1 \cdots p_n$. We next compute $g_1 = \gcd(g, p_1 \cdots p_{\lfloor n/2 \rfloor})$ and $g_2 = \gcd(g, p_{\lfloor n/2 \rfloor+1} \cdots p_n)$. If $g_1 \neq 1$, then we go on with the computation of $g_{1,1} = \gcd(g_1, p_1 \cdots p_{\lfloor n/4 \rfloor})$ and $g_{1,2} = \gcd(g_1, p_{\lfloor n/4 \rfloor+1} \cdots p_{\lfloor n/2 \rfloor})$, and similarly for $g_2$. We repeat this dichotomic process until we have found all prime factors of $q$. For each of the $s$ prime factors $p_j$ of $q$, we next compute $p_j^{k_j} = \gcd(q, p_j^d)$ and $k_j = \log p_j^{k_j} / \log p_j$. The total running time of this algorithm is bounded by $O(n \log^3 n + s \log \Pi \log \log \Pi) = O^\flat(n + s \log \Pi)$. □

THEOREM 6. *Assume the heuristics* **HE** *and* **HC***. Let $\tau > \tau_{\text{crit}}$ with $\tau \leqslant 1$, $N = t_P t_Q |P| |Q|$, and $\Pi \asymp p_n^d T$. Then Algorithm 2 is correct and runs in time*

$$6 \tau \mathsf{M}'_\Pi(T) + O^\flat((s_P + s_Q + s_R) \log \Pi + (t_P + t_Q) \log N + t_R n).$$

**Proof.** We have already explained why our algorithm returns the correct answer with probability $1 - O(\varepsilon)$. The complexity analysis for steps 6 and 8b is similar as the one for Algorithm 1. The running time for the other steps is as follows:

- The reductions of $P$ and $Q$ modulo $\Pi$ can be computed in time $O^\flat((t_P + t_Q) \log N)$, in step 3.
- In step 4, we have to compute $p_i^e$ modulo $\Pi$ for every power $x_i^e$ occurring in the representation of $P$ or $Q$. Using binary powering, such a power can be computed in time $O^\flat(\log d \log \Pi) = O^\flat(\log \Pi)$. The total cost of this step is therefore bounded by $O^\flat((s_P + s_Q) \log \Pi)$.
- Step 5 takes time $O((s_P + s_Q) \log r + (t_P + t_Q) \log \Pi) = O((s_P + s_Q) \log \Pi)$.
- In steps 7 and 8c, we have already shown that one factorization can be done in time $O^\flat(n + s \log \Pi)$. Altogether, the cost of these steps is therefore bounded by $O^\flat(t_R n + s_R \log \Pi)$. □

**Note.** We assumed that $\Pi$ is known in the algorithm. The computation of $\Pi$ can be done in expected time $\tilde{O}(\log^7 \Pi)$: we keep picking $\Pi$ uniformly at random between $\lceil p_n^d T / \varepsilon \rceil$ and $2 \lceil p_n^d T / \varepsilon \rceil$ until we find a prime. This requires an expected number of $O(\log \Pi / \log \log \Pi)$ attemps of cost $\tilde{O}(\log^6 \Pi) = O^\flat(\log^6 \Pi)$ each [36]. For practical purposes, one may use a table of known Mersenne primes instead.

# 5. VARIANTS AND FURTHER EXPERIMENTS

## 5.1. Supersparse polynomials

In section 4.2, we have described an efficient algorithm for the case when the total degree $d$ is small. This indeed holds for most practical applications, but it attractive to also study the case when our polynomials are "truly sparse" with potentially large exponents. This is already interesting for univariate polynomials, a case that we also did not consider so far. Modulo the technique of Kronecker substitution [24, Section 7.1], it actually suffices to consider the univariate case. For simplicity, let us restrict ourselves to this case. When $p_n^d$ gets moderately large, it can also be interesting to consider the incremental approach from [22] and [24, Section 7.3].

One first observation in the univariate case is that we can no longer take the same number of boxes $r$ for our three throws. Next best is to take $r_i = 2\lceil \tau t/2 \rceil + i$ boxes for our $i$-th throw, where $i = 1, 2, 3$. By construction this ensures that $r_1$, $r_2$, and $r_3$ are pairwise coprime. If the exponents of $R$ are large, then it is also reasonable to assume that our heuristic **HE** still holds.

For the efficient determination of the exponents, let $\Pi \geqslant d\,T/\varepsilon$. As before, we may take $\Pi$ to be prime or a product of prime numbers that fit into machine words. Following Huang [29], we now evaluate both $\bar{R} = \bar{P}\,\bar{Q}$ and $x\,\bar{R}' = x\,(\bar{P}'\,\bar{Q} + \bar{P}\,\bar{Q}')$ at $u$ in $(\mathbb{Z}/\Pi\,\mathbb{Z})[u]/(u^{r_i} - 1)$ for $i = 1, 2, 3$. Any term $c\,x^e$ of $\bar{R}$ then gives rise to a term $\tilde{c}\,x^e$ of $x\,\bar{R}'$ with $\tilde{c} = e\,c$, so we can directly read off $e$ from the quotient $\tilde{c}/c$. Modulo the above changes, this allows us to proceed as in Algorithm 2. With the notation

$$f = O^{\approx}(g) \iff f = O(g\,(n \log\,(d\,s_P\,s_Q\,s_R\,|P|\,|Q|\,|R|))^{o(1)}),$$

one may prove in a similar way as before that the bit complexity of determining the exponents in this way is bounded by

$$9\,\tau\,\mathsf{M}'_{\Pi}(T) + O^{\approx}((t_P + t_Q + t_R)\log\,(d\,N\,t_R)),$$

since $s_P \leqslant t_P$, $s_Q \leqslant t_Q$, and $s_R \leqslant t_R$ for univariate polynomials. Returning to the multivariate setting via Kronecker substitution, this yields the bound

$$9\,\tau\,\mathsf{M}'_{\Pi}(T) + O^{\approx}((t_P + t_Q + t_R)\,(n \log d + \log\,(N\,t_R))), \tag{5}$$

where $\log \Pi = O(n \log d + \log T)$. Once the exponents are known, we may use Algorithm 1 to compute the coefficients. In terms of $O^{\approx}$, the complexity becomes

$$3\,\tau\,\mathsf{M}_N(t_R) + O^{\approx}((s_P + s_Q + s_R)\log\,(d\,t_R) + (t_P + t_Q + t_R)\log N). \tag{6}$$

## 5.2. Other rings of coefficients

**Rational coefficients.** Using the combination of modular reduction and rational number reconstruction [12, Chapter 5], the new algorithms generalize in a standard way to polynomials with rational coefficients.

**Complex numbers.** Given a fixed working precision $p$, it is also possible to use coefficients in $\mathbb{C}$, while using a fixed point number representation. Provided that all coefficients have the same order of magnitude, the floating point case can be reduced to this case. Indeed, the reductions modulo $u^r - 1$ are remarkably stable from a numerical point of view, and we may use FFT-multiplication in $\mathbb{C}[u]/(u^r - 1)$. Algorithm 2 can also be adapted, provided that $2^p \gg \lceil B\,T/\varepsilon \rceil$.

**Finite fields.** Our algorithms adapt in a straightforward way to coefficients in a finite field of sufficiently large characteristic. If the characteristic is moderately large (e.g. of the size of a machine word), then one may keep $\Pi$ small in Algorithm 2 by using the incremental technique from [22] and [24, Section 7.3]. Over finite fields $\mathbb{F}_q$ of small characteristic, one needs to use other techniques from [24].

Instead of taking our three additional evaluation points of the form $(p_1 u^{\lambda_{i,1}},\ldots,p_n u^{\lambda_{i,n}})$, we may for instance take them of the form $\big(\omega u^{\lambda_{i,1}},\omega^{d+1} u^{\lambda_{i,2}},\ldots,\omega^{(d+1)^{n-1}} u^{\lambda_{i,n}}\big)$, where $\omega$ is a primitive root of unity of large smooth order (this may force us to work over an extension field; alternatively, one may use the aforementioned incremental technique and roots $\omega$ of lower orders). Thanks to the smoothness assumption, discrete logarithms of powers of $\omega$ can be computed efficiently using Pohlig–Hellman's algorithm. This allows us to recover exponents $k_1 \leqslant d,\ldots,k_n \leqslant d$ from quotients of the form $\tilde{c}/c=\omega^{k_1+k_2(d+1)+\cdots+k_n(d+1)^{n-1}}$.

**General rings.** Finally, it is possible to use coefficients in a general ring $\mathbb{A}$, in which case complexities are measured in terms of the number of operations in $\mathbb{A}$. For rings of sufficiently large characteristic, one may directly adapt the algorithms from section 4. For rings of small characteristic, it is possible to generalize the techniques from the finite field case.

## 5.3. Other operations

The techniques from this paper can also be used for general purpose sparse interpolation. In that case the polynomial $R \in \mathbb{A}[x_1,\ldots,x_n]$ is given through an arbitrary blackbox function that can be evaluated at points in $\mathbb{A}$-algebras over some ring $\mathbb{A}$. This problem was studied in detail in [24]. In section 6, we investigated in particular how to replace expensive evaluations in cyclic $\mathbb{A}$-algebras of the form $\mathbb{A}[x]/(x^r-1)$ by cheaper evaluations at suitable $r$-th roots of unity in $\mathbb{A}$ or a small extension of $\mathbb{A}$. The main new techniques from this paper were also anticipated in section 6.6.

The algorithms from this paper become competitive with the geometric sequence approach if the blackbox function is particularly cheap to evaluate. Typically, the cost $L$ of one evaluation should be $L=O((\log t_R)^2)$ or $L=O((\log t_R)^3)$, depending on the specific variant of the geometric sequence approach that we use.

Technically speaking, it is interesting to note that the problem from this paper actually does *not* fit into this framework. Indeed, if $P$ and $Q$ are sparse polynomial that are represented in their standard expanded form, then the evaluation of $R=PQ$ at a single point requires $L=O(s_P+s_Q+1)$ operations in $\mathbb{A}$, and we typically have $L=\Theta(\sqrt{t_R})$. This is due to the fact that the standard blackbox model does not take into account possible speed-ups if we evaluate our function at many points in geometric progression or at $u$ in a cyclic algebra $\mathbb{A}[u]/(u^r-1)$.

Both in theory and for practical applications, it would be better to extend the blackbox model by allowing for polynomials $R$ of the form

$$R=f(x_1,\ldots,x_n,U_1(x_1,\ldots,x_n),\ldots,U_m(x_1,\ldots,x_n)),$$

where $f \in \mathbb{A}[x_1,\ldots,x_n,y_1,\ldots,y_m]$ is a blackbox function and $U_1,\ldots,U_m \in \mathbb{A}[x_1,\ldots,x_n]$ are sparse polynomials in their usual expanded representation. Within this model, the techniques from this paper should become efficient for many other useful operations on sparse polynomials, such as the computation of gcds or determinants of matrices whose entries are large sparse polynomials.

| $n$ | 2 | 2 | 2 | 3 | 3 | 3 | 4 | 4 | 5 | 7 | 10 |
|---|---|---|---|---|---|---|---|---|---|---|---|
| $d$ | 100 | 250 | 1000 | 25 | 50 | 100 | 20 | 40 | 20 | 15 | 10 |
| $t$ | 5151 | 31626 | 501501 | 3276 | 23426 | 176853 | 10626 | 135751 | 53130 | 170544 | 184756 |
| $\tau$ | 1.14 | 1.14 | 1.14 | 1.14 | 1.14 | 1.14 | 1.11 | 1.14 | 1.14 | 1.17 | 1.20 |

**Table 5.** The particular case when our sparse polynomials $P, Q, R$ are dense polynomials in $n$ variables of given total degrees $d_P$, $d_Q$ and $d = d_R = d_P + d_Q$. The table shows values of $\tau$ for which it is possible to win the game of mystery balls for suitable evaluation points (we do not claim these values to be optimal).

## 5.4. Special types of support

The heuristic **HE** plays an important role in our complexity analysis. It is an interesting question whether it is satisfied for polynomials $R$ whose support is highly regular and not random at all. A particularly important case is when $P, Q$ and $R$ are dense polynomials in $n$ variables of total degrees $d_P$, $d_Q$ and $d = d_R = d_P + d_Q$. Such polynomials are often considered to be sparse, due to the fact that $R$ contains about $n!$ times less terms than a fully dense polynomial of degree $d$ in each variable.

If $R$ is bivariate or trivariate and of very high degree, then it has been shown in [18] that $R$ can be computed in approximately the same time as the product of two dense univariate polynomials which has the same number of terms as $R$. An algorithm for arbitrary dimensions $n$ has been presented in [26], whose cost is approximately $3 - 2/n$ times larger than the cost of a univariate product of the same size. The problem has also been studied in [20, 21] and it is often used as a benchmark [15, 37].

What about the techniques from this paper? We first note that the exponents of $R$ are known, so we can directly focus on the computation of the coefficients. In case that we need to do several product computations for the same $n$ and $d$, we can also spend some time on computing a small $\tau$ and vectors $\lambda_1, \lambda_2, \lambda_3 \in \mathbb{Z}^n$ for which we know beforehand that we will win our game of mystery balls. For our simulations, we found it useful to chose each $\lambda_i$ among a dozen random vectors in a way that minimizes the sum of the squares of the number of balls in the boxes (this sum equals 22 for the first throw in Figure 1).

For various $n$, $d$, and $\tau$, we played our game several times for different triples of vectors $(\lambda_1, \lambda_2, \lambda_3)$. The critical value for $\tau$ for this specific type of support seems to be close to 1.14. In Table 5 below, we report several cases for which we managed to win for this value of $\tau$. This *proves* that optimal polynomial multiplication algorithms for supports of this type are almost as efficient as dense univariate polynomial products of the same size.

# 6. RELEASING THE HEURISTICS

The main results of this paper rely on two heuristics **HE** and **HC**. The first heuristic is the most important one and it is in particular satisfied when $P$ and $Q$ are random sparse polynomials: indeed, in this case, the distributions are equivalent to those obtained by

**Algorithm 3**
**Input:** $P, Q \in \mathbb{Z}[x_1, \ldots, x_n]$ and $e_1, \ldots, e_t \in \mathbb{N}^n$ with $PQ \in \mathbb{Z} x^{e_1} + \cdots + \mathbb{Z} x^{e_t}$
**Output:** the product $PQ$

1. Initialize $\mathcal{I} := \{1, \ldots, t\}$ and $R := 0$.
2. While $\mathcal{I} \neq \emptyset$ do
   a. Let $r$ be a random prime number with $4|\mathcal{I}| < r < 8|\mathcal{I}|$.
   b. Let $\lambda$ be a random element of $\{0, \ldots, r-1\}^n$.
   c. Let $\bar{P} := P(t^{\lambda_1}, \ldots, t^{\lambda_n}), \bar{Q} := Q(t^{\lambda_1}, \ldots, t^{\lambda_n}), \bar{R} := R(t^{\lambda_1}, \ldots, t^{\lambda_n}) \in \mathbb{Z}[t]/(t^r - 1)$.
   d. Compute $\bar{\Delta} := \bar{P}\bar{Q} - \bar{R}$ and write $\bar{\Delta} = \sum_{i \in \mathcal{I}} \delta_i t^{\lambda \cdot e_i}$.
   e. Let $\mathcal{J} := \{i \in \mathcal{I} : \forall j \in \mathcal{I}, \lambda \cdot e_i = \lambda \cdot e_j \Rightarrow e_i = e_j\}$ and $\Delta := \sum_{i \in \mathcal{J}} \delta_i t^{e_i}$.
   f. Set $\mathcal{I} := \mathcal{I} \setminus \mathcal{J}$ and $R := R + \Delta$
3. Return $R$.

fixing linearly independent $\lambda_1$, $\lambda_2$, and $\lambda_3$ and then picking monomials at random. In the case of special supports, the experiments in section 5.4 indicate that the complexity of our algorithms might be even better than for random supports. However, the theoretical question remains what can be proved if we reject the heuristics **HE** and **HC**.

Without **HE** and **HC**, it is easier to analyze the probabilistic cost for repeated single throws with a larger number of boxes $r$. Instead of obtaining the full result from three simultaneous throws by playing our game of mystery balls, we obtain increasingly good approximations of the result as we keep on throwing our balls; see [1, sections 5, 6, and 7] and [24, section 3.1] for this idea. In this section, we present complexity analyses for multiplication algorithms that are based on this approach.

## 6.1. Determination of the coefficients

As in section 4.1, we start with the case when we already know the exponents of the product $PQ$. In that case, we can adapt the number of boxes $r$ at every throw to the number of unknown coefficients, which also simplifies the probabilistic analysis. As before, the resulting Algorithm 3 is Las Vegas.

THEOREM 7. *Let $N = t_P t_Q |P| |Q|$. Then Algorithm 3 is correct and runs in expected time*

$$O(\mathsf{M}_N(t_R)) + O^{\flat}((s_P + s_Q + s_R) \log^2 t_R + (t_P + t_Q + t_R) \log N \log t_R + \log^8 t_R).$$

**Proof.** Let $PQ = c_1 x^{e_1} + \cdots + c_t x^{e_t}$. We claim that $R = \sum_{i \notin \mathcal{I}} c_i x^{e_i}$ at the start (and at the end) of the main loop. This is clearly the case when entering the loop for the first time. It follows that $\bar{\Delta} = \sum_{i \in \mathcal{I}} c_i x^{\lambda \cdot e_i}$. The decomposition $\bar{\Delta} = \sum_{i \in \mathcal{I}} \delta_i t^{\lambda \cdot e_i}$ is not unique, but the coefficient $\delta_i$ is uniquely determined whenever $i \in \mathcal{J}$, in which case $\delta_i = c_i$. This ensures that $\Delta = \sum_{i \in \mathcal{J}} c_i x^{e_i}$ and that we still have $R = \sum_{i \notin \mathcal{I}} c_i x^{e_i}$ at the end of the loop. If the algorithm terminates, then it follows that it returns the correct result.

Let us now show that $\mathcal{J}$ contains at least $|\mathcal{I}|/2$ elements with probability $\geqslant 1/2$. For each pair $(i,j)\in\mathcal{I}^2$ with $i<j$, let $\Lambda_{i,j}$ be the set of $\lambda\in\{0,\ldots,r-1\}^n$ for which $(e_i-e_j)\cdot\lambda=0$. Since $e_i\neq e_j$ modulo $r$, the set $\Lambda_{i,j}$ forms a hyperplane modulo $r$, whence $|\Lambda_{i,j}|=r^{n-1}$. For each $\lambda\in\{0,\ldots,r-1\}^n$, let $P_\lambda$ be the set of pairs $(i,j)\in\mathcal{I}^2$ with $i<j$ and $(e_i-e_j)\cdot\lambda=0$. Then

$$\sum_{\lambda\in\{0,\ldots,r-1\}^n}|P_\lambda| = \sum_{i<j}|\Lambda_{i,j}| = \binom{|\mathcal{I}|}{2}r^{n-1},$$

so the average size of $P_\lambda$ is given by

$$|P_\lambda|_{\mathrm{av}} := \frac{\sum_{\lambda\in\{0,\ldots,r-1\}^n}|P_\lambda|}{|\{0,\ldots,r-1\}^n|} = \binom{|\mathcal{I}|}{2}/r \leqslant \frac{|\mathcal{I}|^2}{2r}.$$

Now for any $\kappa\geqslant 1$, the probability that $|P_\lambda|>\kappa|P_\lambda|_{\mathrm{av}}$ for a random $\lambda$ is bounded by $\Pr[|P_\lambda|>\kappa|P_\lambda|_{\mathrm{av}}]\leqslant\kappa^{-1}$. In particular, we have $|P_\lambda|\leqslant|\mathcal{I}|^2/r$ with probability $\geqslant 1/2$. Since we took $r\geqslant 4|\mathcal{I}|$, it follows that $|P_\lambda|\leqslant|\mathcal{I}|/4$ and thus $|\mathcal{J}|\geqslant|\mathcal{I}|/2$, for such $\lambda$.

Let us next examine the cost of one iteration of the main loop. We can compute $r$ in expected time $O^\flat(\log^7 t_R)$, as in the note after Theorem 6. As in the proof of Theorem 4, the computation of $\bar{P}$, $\bar{Q}$, and $\bar{R}$ requires $O^\flat((s_P+s_Q+s_R)\log t_R+(t_P+t_Q+t_R)\log N)$ bit operations. The multiplication $\bar{P}\bar{Q}$ can be done in time $O(\mathsf{M}_N(r))$. In combination with the above bound, the expected cost to halve the size of $|\mathcal{I}|$ is therefore bounded by $O(\mathsf{M}_N(r))+O^\flat((s_P+s_Q+s_R)\log t_R+(t_P+t_Q+t_R)\log N+\log^7 t_R)$. Using that $r=O(|\mathcal{I}|)$, we conclude that the total expected cost is bounded by $O(\mathsf{M}_N(t_R))+O^\flat((s_P+s_Q+s_R)\log^2 t_R+(t_P+t_Q+t_R)\log N\log t_R+\log^8 t_R)$. □

**Remark 8.** With respect to the bound from Theorem 4, we see that the cost $O(\mathsf{M}_N(t_R))$ of the cyclic polynomial multiplications is just a constant time worse. However, the subdominant terms $O^\flat((s_P+s_Q+s_R)\log t_R+(t_P+t_Q+t_R)\log N)$ in the complexity of Theorem 4 need to be multiplied by $O(\log t_R)$. Whereas $(t_P+t_Q+t_R)\log N\log t_R$ remains bounded by $O(\mathsf{M}_N(t_R))$ for the best known multiplication algorithm with $\mathsf{M}_N(t_R)\asymp t_R\log N\log(t_R\log N)$, the term $(s_P+s_Q+s_R)\log^2 t_R$ may dominate the cost if $s_R\asymp n\,t_R$ and $\log N$ is small in comparison with $n$.

**Remark 9.** In this paper, we focus on sparse polynomial multiplication, but the approach from Algorithm 3 can also be applied to the sparse interpolation of a polynomial $R\in\mathbb{Z}[x_1,\ldots,x_n]$ whose exponents are known, by adapting the algorithms from [24].

## 6.2. Determination of the exponents

Let us now examine the case when the exponents are not known. Then we in particular have to guess the number of terms of the product $PQ$. We do this by introducing a "tentative upper bound $T$" for the number of "missing terms" $t_{PQ-R}$, which is updated along with the approximation $R$ of $PQ$. We start with $T=4$ and, as long as $T<t_{PQ-R}$, we will show that $T$ doubles with high probability at every iteration.

In order to avoid depending on the heuristic **HC**, we also compute an extra polynomial $P_3$ in step 5 of Algorithm 4. This allows us to replace the heuristic by a probabilistic consistency check in step 9c. This only multiplies the complexity by a constant factor. Note that the same trick can be used in combination with Algorithm 2.

In order to ease the probabilistic analysis, we first consider the case when the polynomials $P$ and $Q$ have modular coefficients in $\mathbb{Z}/\Pi\,\mathbb{Z}$.

**Algorithm 4**
**Input:** $P, Q \in (\mathbb{Z}/\Pi\mathbb{Z})[x_1, \ldots, x_n]$, where $\Pi \geqslant 4 p_n^{d_{PQ}}$ is prime
**Output:** the product $PQ$, with high probability (Monte Carlo)

1. Initialize $R := 0$ and $T := 8$
2. Let $r$ be a random prime number with $8T < r < 16T$.
3. Let $\lambda$ be a random element of $\{0, \ldots, r-1\}^n$.
4. Let $v_1, \ldots, v_n, w_1, \ldots, w_n$ be random elements of $(\mathbb{Z}/\Pi\mathbb{Z})^*$.
5. Let $P_1 := P(v_1 x_1, \ldots, v_n x_n)$, $Q_1 := Q(v_1 x_1, \ldots, v_n x_n)$, $R_1 := R(v_1 x_1, \ldots, v_n x_n)$.
   Let $P_2 := P_1(p_1 x_1, \ldots, p_n x_n)$, $Q_2 := Q_1(p_1 x_1, \ldots, p_n x_n)$, $R_2 := R_1(p_1 x_1, \ldots, p_n x_n)$.
   Let $P_3 := P_1(w_1 x_1, \ldots, w_n x_n)$, $Q_3 := Q_1(w_1 x_1, \ldots, w_n x_n)$, $R_3 := R_1(w_1 x_1, \ldots, w_n x_n)$.
6. Let $\bar{P}_\ell := P_\ell(t^{\lambda_1}, \ldots, t^{\lambda_n}), \bar{Q}_\ell := Q_\ell(t^{\lambda_1}, \ldots, t^{\lambda_n}), \bar{R}_\ell := R_\ell(t^{\lambda_1}, \ldots, t^{\lambda_n}) \in (\mathbb{Z}/\Pi\mathbb{Z})[t]/(t^r - 1)$, for $\ell = 1, 2, 3$.
7. Compute $\bar{\Delta}_\ell := \bar{P}_\ell \bar{Q}_\ell - \bar{R}_\ell$ for $\ell = 1, 2, 3$ and write $\bar{\Delta}_\ell = \sum_{j<r} \bar{\Delta}_{\ell,j} t^j$.
8. If $\bar{\Delta}_1 = \bar{\Delta}_2 = \bar{\Delta}_3 = 0$, then return $R$.
9. For $j = 0, \ldots, r-1$ with $\bar{\Delta}_{1,j} \neq 0$ do:
   a. Compute the preimage $q \in \{0, \ldots, \Pi - 1\}$ of $\bar{\Delta}_{2,j} / \bar{\Delta}_{1,j}$.
   b. If $q \leqslant p_n^{d_{PQ}}$, then try factoring $q = p_1^{k_1} \cdots p_n^{k_n}$.
   c. In case of success, check whether $\bar{\Delta}_{3,j} = \bar{\Delta}_{1,j} w_1^{k_1} \cdots w_n^{k_n}$.
   d. In case of resuccess, set $R := R + \bar{\Delta}_{1,j} (x_1/v_1)^{k_1} \cdots (x_n/v_n)^{k_n}$
10. Let $\mathcal{E}$ be the set of indices $j$ with $\bar{\Delta}_{1,j} \neq 0$ for which one of the above checks failed.
11. Set $T := \max(16|\mathcal{E}|, T/2)$ and jump to step 2.

THEOREM 10. *Let $\eta := d_{PQ}/(\Pi - 1)$, $\varepsilon := 1 - (1 - n\eta)^{t_{PQ}}$, and assume that $\eta \leqslant \varepsilon \leqslant 1/2$. Then with probability at least $1 - \varepsilon - \eta$, the Algorithm 4 returns the correct answer and runs in expected time*

$$O(\mathsf{M}_\Pi(t_R)) + O^\flat((s_P + s_Q + s_R) \log \Pi \log t_R + (t_P + t_Q + t_R) \log^2 t_R + t_R n + \log^8 t_R). \qquad (7)$$

**Proof.** We will say that an execution of the algorithm is *flawless* if, throughout the execution, every term $c x^e$ that occurs in $R$ also occurs in $PQ$. This is the case if we only add correct terms to $R$ in step 9d. Let us analyze the probability that this is indeed the case.

Assume that we arrive at step 9 and that the execution has been flawless until that point. For a given $j$, let $c_1 x^{e_1}, \ldots, c_\ell x^{e_\ell}$ be all terms that occur in $\Delta_1$ with $e_i \cdot \lambda = j$ modulo $r$. Setting $A(x) := c_1 x^{e_1} + \cdots + c_\ell x^{e_\ell}$ and $B(x) = A(x) - A(1)\, x_1^{k_1} \cdots x_n^{k_n}$, we note that $B(w) = \bar{\Delta}_{3,j} - \bar{\Delta}_{1,j} w_1^{k_1} \cdots w_n^{k_n}$. Since the execution was flawless until here, we have $d_A \leqslant d_{PQ}$ and $d_B \leqslant n\, d_{PQ}$. If $\ell > 1$, then the Schwarz–Zippel lemma [44, 46] implies that the probability that we pass the check in step 9c is bounded by $d_B/(\Pi - 1) \leqslant n\eta$. The probability that the term $\bar{\Delta}_{1,j} (x_1/v_1)^{k_1} \cdots (x_n/v_n)^{k_n}$ in step 9d is correct is therefore at least $1 - n\eta$ and the probability that this is the case for all $t_{PQ}$ terms of $PQ$ is at least $(1 - n\eta)^{t_{PQ}} = 1 - \varepsilon$. We conclude that the probability that the execution is flawless is at least $1 - \varepsilon$.

Assume next that a flawless execution terminates and let us examine the probability that the returned answer is incorrect. So assume that $\Delta := PQ - R \neq 0$ and let $c x^e$ be one of the terms occurring in $\Delta$. Let $c_1 x^{e_1}, \ldots, c_\ell x^{e_\ell}$ be all terms that occur in $\Delta$ with $e_i \cdot \lambda = e \cdot \lambda$ modulo $r$. Let $A(x) := c_1 x^{e_1} + \cdots + c_\ell x^{e_\ell}$ and note that $\bar{\Delta}_{1,e\cdot\lambda} = A(v)$. Now $d_A \leqslant d_{PQ}$ by our flawless assumption. By the Schwarz–Zippel lemma, it follows that the probability that $A(v) = 0$ is at most $d_A / (\Pi - 1) \leqslant d_{PQ} / (\Pi - 1)$. This shows that the probability that the algorithm returns an incorrect answer is bounded by $d_A / (\Pi - 1) \leqslant \eta$. Since the last random choices of the $v_i$ and $w_i$ are independent from the previous ones that led to the flawless execution, the overall probability of success is at least $(1 - \varepsilon)(1 - \eta) \geqslant 1 - \varepsilon - \eta$.

As to the complexity bound, let us again focus on a flawless execution. Note that the flawless assumption only involves the random choices of the $v_i$ and $w_i$ in step 4; the $\lambda_i$ in step 3 are chosen independently; we will freely use this observation in the probabilistic analysis below. Let us examine the evolution of $T$ and $t := t_{PQ-R}$ during a flawless execution. Clearly, $t$ can only decrease. Let $T' := \max(16|\mathcal{E}|, T/2)$ and $t'$ be the new values of $T$ and $t$ after one iteration of the main loop (which starts at step 2 and ends at step 11). If $T$ is an upper bound for $t$, then by similar arguments as in the proof of Theorem 7, we have $t' \leqslant t/2$ with probability $\geqslant 3/4$ (note that we now took $r \geqslant 8T$ instead of $r \geqslant 4T$).

If $T < t$ and $t' > t/2$, then we claim that $T' \geqslant 2T$ with probability $\geqslant 3/4$. Assume that $T' \leqslant 2T$, whence $T' \leqslant 2T < 2t < 4t'$. Let $P_\lambda$ be the set of pairs $(x^e, x^{e'})$ of monomials in $PQ - R$ with $e < e'$ (for any total order on the exponents) and whose images in $(\mathbb{Z}/\Pi\mathbb{Z})[t]/(t^r - 1)$ under the map $x^e \mapsto t^{\lambda \cdot e}$ are the same. As in the proof of Theorem 7, we have $|P_\lambda|_{\mathrm{av}} \leqslant t^2/(2r)$. Now the image of a pair $(x^e, x^{e'})$ in $P_\lambda$ is of the form $(t^j, t^j)$ with $j \in \mathcal{E}$. For a given $\mathcal{E}$, the minimum of $|P_\lambda|$ is reached when the monomials are "equidistributed" over the "boxes" in $\mathcal{E}$. In that case, every $t^j$ with $j \in \mathcal{E}$ has at least $\lfloor t'/|\mathcal{E}| \rfloor$ pre-images among the monomials in $P_\lambda$. Then it follows that

$$
\begin{aligned}
|P_\lambda| &\geqslant |\mathcal{E}| \binom{\lfloor t'/|\mathcal{E}| \rfloor}{2} \\
&\geqslant \frac{1}{2|\mathcal{E}|}(t' - 2|\mathcal{E}|)(t' - |\mathcal{E}|) \\
&\geqslant \frac{16}{2T'}\left(t' - \frac{2T'}{16}\right)\left(t' - \frac{T'}{16}\right) \\
&= 8\left(1 - \frac{T'}{8t'}\right)\left(1 - \frac{T'}{16t'}\right)\frac{(t')^2}{T'} \\
&\geqslant 8\left(1 - \frac{1}{2}\right)\left(1 - \frac{1}{4}\right)\frac{(t')^2}{T'} \\
&= 3\frac{(t')^2}{T'},
\end{aligned}
$$

whereas

$$
|P_\lambda|_{\mathrm{av}} \leqslant \frac{t^2}{2r} < \frac{t}{16T}t \leqslant \frac{1}{2} \cdot \frac{(t')^2}{T'}.
$$

Consequently, $|P_\lambda| > 4|P_\lambda|_{\mathrm{av}}$, which can only happen with probability $\leqslant 1/4$. This completes the proof of our claim. Our claim implies that after an average $O(1)$ number of iterations, we reach a situation in which either $T$ is doubled or $t$ is halved.

Now assume that $t \leqslant T$ at a certain point of the execution. Then with probability $\geqslant 3/4$, we have $t' \leqslant t/2 \leqslant T/2 \leqslant T'$ after one iteration. With probability $\leqslant 1/4$, we need an average number of $O(1)$ additional iterations before $t$ is either halved or we again reach a point when $t \leqslant T$ (here we use the fact that we always have $T' \geqslant T/2$ after one itera-tion). Combining these two cases, we see that after an average number of $O(1)$ iterations, the number $t$ of missing terms is halved and $T$ is again an upper bound for $t$. Using a similar analysis as in the proof of Theorem 7, it follows that the overall time before we terminate (where we count from the first moment when $t \leqslant T$ holds) is bounded by (7).

It remains to estimate the cost of the initial phase of reaching a point when $t \leqslant T$ holds. By our claim, we need an average number of $O(1)$ iterations in order to double $T$. The cost of these iterations is bounded by $O(\mathsf{M}_\Pi(T)) + O^b((s_P + s_Q + s_R) \log \Pi + (t_P + t_Q + t_R) \log t_R + T n)$, as in the proofs of Theorems 6 and 7. Summing this bound for $T$ in a geometric progression until $T \geqslant t$, the cost of the initial phase is therefore again bounded by (7). □

*Mutatis mutandis*, we note that Remarks 8 and 9 also apply for Theorem 10 and Algorithm 4. It remains to consider the original problem of determining the exponents of $R = P Q$ in the case when $P$ and $Q$ have coefficients in $\mathbb{Z}$. For this it suffices to examine the probability that a random modular reduction of $R$ has the same exponents as $R$.

PROPOSITION 11. *Let $R \in \mathbb{Z}[x_1, \ldots, x_n]$ be a polynomial with $t$ terms and $|P| \leqslant 2^B$. Let $b \geqslant 30$ with $2^b \geqslant 10 t B$ and let $\Pi$ be a random prime number between $2^b$ and $2^{b+1}$. Then the probability that $R$ and $R \operatorname{mod} \Pi$ have the same exponents is at least $1 - 5 t B / 2^b$.*

**Proof.** The $t$ coefficients of $R$ are divisible by at most $(t B) / b$ distinct primes between $2^b$ and $2^{b+1}$. Let $\pi(x)$ be the number of primes below $x$. By [43], we have

$$\frac{x}{\log x + 2} < \pi(x) < \frac{x}{\log x - 4}$$

for $x \geqslant 55$. Consequently,

$$\pi(2^{b+1}) - \pi(2^b) \geqslant \frac{2^b}{b \log 2} \left( \frac{2}{1 + \frac{\log 2 + 2}{b \log 2}} - \frac{1}{1 - \frac{4}{b \log 2}} \right) \geqslant \frac{1}{2} \cdot \frac{2^b}{b \log 2}$$

for $b \geqslant 30$. The probability that a randomly chosen prime between $2^b$ and $2^{b+1}$ does not divide any of the coefficients of $R$ is therefore at least $(1 - N^{-1})^{(tB)/b}$ where $N = 2^{b-1} / \log 2^b$. Under our assumption that $2^b \geqslant 10 t B$, we have $(1 - N^{-1})^{(tB)/b} \geqslant 1 - 5 t B / 2^b$. □

Of course, the variants from section 5 also admit unconditional counterparts. In particular, for supersparse polynomials, the bounds (5) and (6) now become

$$O(\mathsf{M}'_\Pi(T)) + O^{\approx}((t_P + t_Q + t_R)\,(n \log d + \log (N t_R)) \log t_R + \log^8 t_R) \qquad \text{resp.}$$
$$O(\mathsf{M}_N(t_R)) + O^{\approx}(((s_P + s_Q + s_R) \log (d t_R) + (t_P + t_Q + t_R) \log N) \log t_R + \log^8 t_R).$$

## BIBLIOGRAPHY

[1] A. Arnold, M. Giesbrecht, and D. S. Roche. Sparse interpolation over finite fields via low-order roots of unity. In *Proc. ISSAC '14*, pages 27–34. New York, NY, USA, 2014. ACM.

[2] A. Arnold, M. Giesbrecht, and D. S. Roche. Faster sparse multivariate polynomial interpolation of straight-line programs. *JSC*, 75:4–24, 2016.

[3] A. Arnold and D. S. Roche. Multivariate sparse interpolation using randomized Kronecker substitutions. In *ISSAC '14: Proceedings of the 39th International Symposium on Symbolic and Algebraic Computation*, pages 35–42. New York, NY, USA, 2014. ACM Press.

[4] M. Asadi, A. Brandt, R. Moir, and M. M. Maza. Sparse polynomial arithmetic with the BPAS library. In *International Workshop on Computer Algebra in Scientific Computing*, pages 32–50. Springer, 2018.

[5] M. Ben-Or and P. Tiwari. A deterministic algorithm for sparse multivariate polynomial interpolation. In *Proc. ACM STOC '88*, pages 301–309. New York, NY, USA, 1988.

[6] J. Canny, E. Kaltofen, and Y. Lakshman. Solving systems of non-linear polynomial equations faster. In *Proceedings of the ACM-SIGSAM 1989 International Symposium on Symbolic and Algebraic Computation*, pages 121–128. ACM Press, 1989.

[7] C. Fieker, W. Hart, T. Hofmann, and F. Johansson. Nemo/Hecke: computer algebra and number theory packages for the Julia programming language. In *Proc. ISSAC 2017*, pages 157–164. 2017.

[8] Ph. Flajolet and R. Sedgewick. *An introduction to the analysis of algorithms*. Addison Wesley, Reading, Massachusetts, 2nd edition, 1996.

[9] T. S. Freeman, G. M. Imirzian, E. Kaltofen, and Y. Lakshman. DAGWOOD: a system for manipulating polynomials given by straight-line programs. *ACM Trans. Math. Software*, 14:218–240, 1988.

[10] S. Garg and É. Schost. Interpolation of polynomials given by straight-line programs. *Theoretical Computer Science*, 410(27-29):2659–2662, 2009.

[11] M. Gastineau and J. Laskar. Development of TRIP: fast sparse multivariate polynomial multiplication using burst tries. In *International Conference on Computational Science*, pages 446–453. Springer, 2006.

[12] J. von zur Gathen and J. Gerhard. *Modern Computer Algebra*. Cambridge University Press, 2-nd edition, 2002.

[13] M. Giesbrecht and D. S. Roche. Diversification improves interpolation. In *ISSAC '11: Proceedings of the 36th International Symposium on Symbolic and Algebraic Computation*, pages 123–130. ACM Press, 2011.

[14] M. T. Goodrich and M. Mitzenmacher. Invertible Bloom lookup tables. In *49th Annual Allerton Conference on Communication, Control, and Computing*, pages 792–799. 2011.

[15] A. W. Groves and D. S. Roche. Sparse polynomials in FLINT. *ACM Commun. Comput. Algebra.*, 50(3):105–108, 2016.

[16] D. Harvey and J. van der Hoeven. Integer multiplication in time $O(n \log n)$. *Annals of Mathematics*, 193(2):563–617, 2021.

[17] H. Hassanieh, P. Indyk, D. Katabi, and E. Price. Nearly optimal sparse Fourier transform. In *Proceedings of the forty-fourth annual ACM symposium on Theory of computing*, pages 563–578. 2012.

[18] J. van der Hoeven. The truncated Fourier transform and applications. In *Proc. ISSAC 2004*, pages 290–296. Univ. of Cantabria, Santander, Spain, July 2004.

[19] J. van der Hoeven. Ball arithmetic. Technical Report, HAL, 2009. https://hal.archives-ouvertes.fr/hal-00432152.

[20] J. van der Hoeven and G. Lecerf. On the complexity of blockwise polynomial multiplication. In *Proc. ISSAC '12*, pages 211–218. Grenoble, France, July 2012.

[21] J. van der Hoeven and G. Lecerf. On the bit-complexity of sparse polynomial multiplication. *JSC*, 50:227–254, 2013.

[22] J. van der Hoeven and G. Lecerf. Sparse polynomial interpolation in practice. *ACM Commun. Comput. Algebra*, 48(3/4):187–191, 2015.

[23] J. van der Hoeven and G. Lecerf. Implementing number theoretic transforms. Technical Report, HAL, 2024. https://hal.science/hal-04841449.

[24] J. van der Hoeven and G. Lecerf. Sparse polynomial interpolation: faster strategies over finite fields. *AAECC*, 2024. https://doi.org/10.1007/s00200-024-00655-5.

[25] J. van der Hoeven et al. GNU TeXmacs. https://www.texmacs.org, 1998.

[26] J. van der Hoeven and É. Schost. Multi-point evaluation in higher dimensions. *AAECC*, 24(1):37–52, 2013.

[27] M. A. Huang and A. J. Rao. Interpolation of sparse multivariate polynomials over large finite fields with applications. In *SODA '96: Proceedings of the seventh annual ACM-SIAM symposium on Discrete algorithms*, pages 508–517. Philadelphia, PA, USA, 1996. Society for Industrial and Applied Mathematics.

[28] Q. L. Huang and X. S. Gao. Sparse Polynomial Interpolation with Finitely Many Values for the Coefficients. In V. Gerdt, V. Koepf, W. Seiler, and E. Vorozhtsov, editors, *Computer Algebra in Scientific Computing. 19th International Workshop, CASC 2017, Beijing, China, September 18-22, 2017, Proceedings.*, volume 10490 of *Lect. Notes Comput. Sci.* Springer, Cham, 2017.

[29] Q.-L. Huang. Sparse polynomial interpolation over fields with large or zero characteristic. In *Proc. ISSAC '19*, pages 219–226. ACM, 2019.

[30] Q.-L. Huang and X.-S. Gao. Revisit sparse polynomial interpolation based on randomized Kronecker substitution. In *Computer Algebra in Scientific Computing, CASC '19*, pages 215–235. Springer, 2019.

[31] M. Javadi and M. Monagan. Parallel sparse polynomial interpolation over finite fields. In *Proceedings of PASCO 2010*, pages 160–168. ACM Press, 2010.

[32] S. C. Johnson. Sparse polynomial arithmetic. *SIGSAM Bull.*, 8(3):63–71, 1974.

[33] E. Kaltofen and L. Yagati. Improved sparse multivariate polynomial interpolation algorithms. In *ISSAC '88: Proceedings of the International Symposium on Symbolic and Algebraic Computation*, pages 467–474. Springer Verlag, 1988.

[34] M. Kapralov. Sparse Fourier transform in any constant dimension with nearly-optimal sample complexity in sublinear time. In *Proceedings of the forty-eighth annual ACM symposium on Theory of Computing*, pages 264–277. 2016.

[35] M. Khochtali, D. S. Roche, and X. Tian. Parallel sparse interpolation using small primes. In *Proceedings of the 2015 International Workshop on Parallel Symbolic Computation*, pages 70–77. 2015.

[36] H. W. Lenstra, Jr. and C. Pomerance. Primality testing with Gaussian periods. https://math.dartmouth.edu/~carlp/PDF/complexity12.pdf, 2005.

[37] M. Monagan and R. Pearce. Parallel sparse polynomial multiplication using heaps. In *ISSAC '09: Proceedings of the 2009 International Symposium on Symbolic and Algebraic Computation*, pages 263–270. ACM Press, 2009.

[38] M. Monagan and R. Pearce. Sparse polynomial multiplication and division in Maple 14. Available from http://cgi.cecm.sfu.ca/~pborwein/MITACS/highlights/SDMPmaple14.pdf, 2009.

[39] H. Murao and T. Fujise. Modular algorithm for sparse multivariate polynomial interpolation and its parallel implementation. *JSC*, 21:377–396, 1996.

[40] C. H. Papadimitriou. *Computational Complexity*. Addison-Wesley, 1994.

[41] D. S. Roche. Chunky and equal-spaced polynomial multiplication. *JSC*, 46(7):791–806, 2011.

[42] D. S. Roche. What can (and can't) we do with sparse polynomials? In *Proc. ISSAC '18*, pages 25–30. New York, NY, USA, 2018. ACM.

[43] B. Rosser. Explicit bounds for some functions of prime numbers. *American Journal of Mathematics*, 63(1):211–232, 1941.

[44] J. T. Schwartz. Fast probabilistic algorithms for verification of polynomial identities. *JACM*, 27(4):701–717, 1980.

[45] T. Yan. The geobucket data structure for polynomials. *JSC*, 25(3):285–293, 1998.

[46] R. Zippel. Probabilistic algorithms for sparse polynomials. In *Proc. EUROSAM '79*, pages 216–226. Springer, 1979.